 \documentclass[prl,twocolumn,showpacs,floats,aps,floatfix]{revtex4}
\usepackage{amsmath}
\usepackage[mathscr]{eucal}
\usepackage{mathrsfs}
\usepackage{pstricks}
\newpsobject{grilla}{psgrid}{subgriddiv=1,griddots=10,gridlabels=6pt}
\newpsobject{ast}{psdot}{dotstyle=asterisk,dotsize=5pt}
\def\ea{{\it et al.}}

\newcommand{\beq}{\begin{equation}}
\newcommand{\eeq}{\end{equation}}
\newcommand{\beqa}{\begin{eqnarray}}
\newcommand{\eeqa}{\end{eqnarray}}
\def\lsim{\ \rlap{\raise 3pt \hbox{$<$}}{\lower 3pt \hbox{$\sim$}}\ }
\def\gsim{\ \rlap{\raise 3pt \hbox{$>$}}{\lower 3pt \hbox{$\sim$}}\ }

\begin{document} 

\title{ 
\preprint{hep-ph/0306206} 
\preprint{UdeA-PE/03-10} 
\large 
 $b$-$\tau$ Yukawa non-unification in supersymmetric $\mathbf{SU(5)}$
 with an Abelian flavor symmetry 
}

\author{Diego Aristizabal${}^1$ and Enrico Nardi${}^{1,2}$} 

\affiliation{ $^1$Departamento de F{\'\i}sica, Universidad de
Antioquia, A.A. {\it 1226} Medell{\'\i}n, Colombia \\
$^2$INFN, Laboratori Nazionali di Frascati, C.P. 13, 00044
Frascati,Italy }

                      \date{\today}    

\begin{abstract} 
\noindent
We present a supersymmetric $SU(5)\times U(1)_H$ model, free from
gauge anomalies, where the Abelian factor $U(1)_H$, introduced to
account for the hierarchy of fermion masses and mixings, is broken by
the same adjoint representations of $SU(5)$ that also break the GUT
symmetry. The model predicts approximate, but never exact, $b$-$\tau$
Yukawa unification. The deviations  are related to group
theoretical coefficients and can naturally be at the level of 10\%-20\%.
\end{abstract} 


\maketitle



\section{Introduction}
 \label{sec:introduction}
\vspace{-4mm}
The Standard Model (SM) of elementary particles represents one of the
greatest achievements in theoretical physics of the past century.
Particle interactions are derived from local symmetries and are
explained at a fundamental level by means of the gauge principle.
However, the SM does not provide any real clue for understanding other
elementary particle properties, like the fermion masses and mixing
angles, that are simply accommodated within the theory.  Among the
several approaches put forth to tackle this problem, the one proposed
long ago by Froggatt and Nielsen (FN) \cite{Froggatt:1978nt} always
attracted much interest in so far as it is able to account
semi-quantitatively for the fermion mass pattern and to yield a couple
of predictions.  The basic ingredient is a horizontal Abelian symmetry
$U(1)_H$ that forbids, at tree level, most of the fermion Yukawa
couplings. The symmetry is spontaneously broken by the vacuum
expectation value (vev) $\langle S\rangle$ of a SM singlet
field. After $U(1)_H$ is broken a set of effective operators arises
that couple the SM fermions to the electroweak Higgs boson, and that
are induced by heavy vectorlike fields with mass $M$.  The hierarchy
of fermion masses results from the dimensional hierarchy among the
various higher order operators, that are suppressed by powers of a
small parameter $\epsilon = \langle S \rangle/M$.  In turn, the
suppression powers are determined by the horizontal charges assigned
to the fermion fields.  In the past, this mechanism was thoroughly
studied in different contexts like the supersymmetric SM
\cite{Leurer:1993gy} or in frameworks where the horizontal symmetry is
promoted to a gauge symmetry that can be anomalous
\cite{Ibanez:1994ig,Binetruy:1994ru,GaugeU1,Nir:1995bu} or
non-anomalous \cite{Mira:1999fx}. In this work we investigate the
consequences of embedding the FN mechanism within a supersymmetric
$SU(5)$ GUT. This appears as a natural step to take, given that the
precise unification of the three gauge couplings within the minimal
supersymmetric SM (MSSM) promoted the GUT idea to an almost compelling
ingredient for a more fundamental theory.

However, a straightforward implementation of the FN mechanism within a
Gran Unified model is not an easy task. Few models have been
constructed in which, differently from our case, an anomalous Abelian
symmetry is used \cite{anomalousGUT}.  Indeed, in the context of GUTs,
non-Abelian flavor symmetries have been often preferred for model
building \cite{nonabelian}. Within a GUT, the main difficulties with
the FN mechanism arise because while in the SM there are five
different multiplets per generation, and correspondingly five
independent horizontal charges, this number is reduced to two in
$SU(5)$, and to a single one in models in which a full fermion
generation is assigned to the same gauge multiplet, like for example
in $SO(10)$.  Since the horizontal charges are free parameters of the
model, it is clear that GUTs symmetries overconstrain the FN
mechanism. Let us just mention the problem represented by mass ratios
like $m_{\mu}/m_{s}\sim m_{d}/m_{e}\sim 3\,$ whose solution has always
been a challenge for GUT model building \cite{Georgi:1979df}.  Due to
the fact that the leptons and the down-type quarks belong to the same
gauge multiplets, it is clear that these mass relations cannot be
explained simply by means of a suitable assignment of the horizontal
charges.

The main aim of this work is to propose a mechanism that seems capable
to account for the problematic mass relations, and therefore
could reconcile the FN approach with the GUT idea. Here we will mainly
focus on the $m_b/m_\tau$ mass ratio, and we will show that in our
framework it could well deviate from unity,  most naturally at the
level of 10\%-20\%.  Such a possibility appears important in view of
the fact that the low energy values of the $b$ and $\tau$ Yukawa
couplings, when run up to the GUT scale by means of the MSSM
renormalization group equations, do not unify with a precision
comparable to gauge coupling unification.  Yukawa unification at an
acceptable level can be achieved only if strong restrictions are
imposed on the MSSM parameter space, and a set of conditions for the
supersymmetric particles masses are satisfied \cite{b-tau}.  This fact
also prompted for investigations of supersymmetric models with
non-universal boundary conditions, that can better accommodate
$b$-$\tau$ unification while satisfying other low energy constraints
\cite{b-tau-nonuniv}.


\vspace{-6mm}

\section{The Model}
 \label{sec:themodel}

\vspace{-4mm}

In $SU(5)$ GUTs, the $SU(2)$ lepton doublets $L$ and the down-quark
singlets $d^c$ are assigned to the fundamental conjugate
representation of the group $\mathbf{\bar 5}$, while quark doublets
$Q$, up-type quark singlets $u^c$ and lepton singlets $e^c$ fill up
the antisymmetric $\mathbf{10}$.  The Higgs field $\phi_d$ responsible
for the down-quarks and lepton masses belong to another $\mathbf{\bar
5}^{\phi_d}$, while $\phi_u$ responsible for the masses of the
$u$-quarks is assigned to a fundamental $\mathbf{5}^{\phi_u}$.
Schematically, the Yukawa Lagrangian reads
\begin{equation}
\label{eq:schem}
 {\cal L}_Y = 
y_{ij}^d\> \mathbf{\bar 5}_i \,
\mathbf{10}_j \> \mathbf{\bar 5}^{\phi_d}+
 y_{ij}^u\> \mathbf{10}_i\, \mathbf{10}_j \,\mathbf{5}^{\phi_u}, 
\end{equation}
where $y^d$ represent the down-quarks and leptons Yukawa couplings,
$y^u $ the up-quark couplings, and $i\,,j=1,2,3$ are generation
indices.

Under the assumption that an additional $U(1)_H$ flavor symmetry is
present and that the fermion masses are generated via the FN
mechanism, the couplings $y_{ij}^{d,u}$ are no more simple
dimensionless numbers, but embed suppression factors that account for
the fact that the two terms in (\ref{eq:schem}) now are effective
operators of the low energy theory, that only arise after the breaking
of $U(1)_H$. By introducing the notation $f_i$ for the horizontal
charge of $\mathbf{\bar 5}_i$, $t_i$ for the charge of $\mathbf{10}_i$
and $f_u$ and $f_d$ for the multiplets containing the up and down-type
Higgs fields, the suppression factors can be written explicitly as:
\begin{equation}\label{explicit}
y_{ij}^d = Y_{ij}^d \> \epsilon^{f_i+t_j+f_d}\,, \qquad
y_{ij}^u = Y_{ij}^u \> \epsilon^{t_i+t_j+f_u}. 
\end{equation}
As mentioned above, $\epsilon\,$ is a small number, with horizontal
charge $-1$, that arises from the ratio between the vev that breaks
$U(1)_H$ and the large mass $M$ of the heavy vectorlike FN fields
needed to generate the effective operators. The
numbers $Y_{ij}^{d,u}$  in (\ref{explicit})
are assumed to be all of order unity, as is
indeed more natural for dimensionless couplings.  Phenomenologically,
the mass hierarchy for the up-type quarks is much stronger than for
the down quarks and leptons. This feature be can easily reproduced
by means of  the following charge assignment:
$t_1=t_2+1=t_3+2\,$ and $f_i=f$ for each of the
three $\mathbf{\bar 5}$. This yields
$m_u\,$:$\>m_c\,$:$\>m_t\approx \epsilon^4\,$:$\> \epsilon^2\,$:$\>1$
and $m_{d,\,e}\,$:$\>m_{s,\,\mu}\,$:$\>m_{b,\,\tau}\approx
\epsilon^2\,$:$\> \epsilon\,$:$\>1$. By choosing $\epsilon \approx
1/25$ the resulting mass ratios are qualitatively correct.  Since the
top Yukawa coupling is of order unity, it must be allowed by the
horizontal symmetry. This implies $2\,t_3+f_u=0$ and suggests the simple
choice $t_3=f_u=0\,$.  The hierarchy $m_{b,\tau}/m_t\ll 1$ can also be
easily accounted for by choosing $f+f_d=t_3+f_u+1=1$.  Let us note
that a redefinition $f\to f-x$ and $f_d\to f_d+x$ with $x$ an
arbitrary number, while it can affect the superpotential Higgs mass
parameter $\mu\,\phi_d\,\phi_u$, it leaves invariant all the Yukawa couplings
in (\ref{explicit}). This is enough freedom to ensure that is it always
possible to set to zero the $SU(5)^2\times U(1)_H$ mixed anomaly
$\propto 3f+\sum_i t_i +f_d+f_u -2x\,$. The pure $U(1)^3_H$ and the mixed
gravitational anomalies can always be canceled by adding SM singlets
with suitable horizontal charges, and therefore $U(1)_H$ can be
straightforwardly gauged, without implying additional constraints on
the horizontal charges.  This is quite different from the SM case
where consistency with phenomenology implies that either $U(1)_{H}$ is
an anomalous symmetry \cite{Binetruy:1994ru,Nir:1995bu} or that the
mass of the up quark must vanish \cite{{Mira:1999fx}}.

It is noticeable how the gross features of the fermion mass hierarchy
can be reproduced by such a simple scheme. Even more interestingly,
when we assume that neutrinos masses are generated from dimension five
operators yielding mass term $m_{ij}^\nu\> \mathbf{\bar
5}_i\,\mathbf{\bar 5}_j\,$, the scheme predicts no hierarchy between
the neutrino masses and unsuppressed mixings.  The last prediction is
in qualitative agreement with the most recent results from neutrino
oscillation experiments, while the first one is certainly a viable
possibility. In fact, this simple scheme has been previously discussed
in relation with anarchical neutrino mass matrices \cite{Haba:2000be},
and it has been shown that it is able to reproduce the known phenomenology in
the neutrino sector (see however the criticisms in
\cite{Espinosa:2003qz}).

Indeed it is somewhat disappointing that at a closer quantitative
inspection the scheme produces too many wrong numerical results.
Besides the two down-quark to leptons mass ratios mentioned above, the
Cabibbo angle $\theta_C$ also turns out to be way too small, simply
because the parameter $\epsilon$ that determines the suppression of
the various mixings is much smaller than $\theta_C$.  In our opinion
these discrepancies are too serious to be accounted for by simply
appealing to random fluctuations in the values of the couplings
$Y_{ij}$.

Nevetheless, as we will now discuss, the scheme can be retained if one
of the initial ingredients of the FN mechanism is modified. While it
is generally assumed that the horizontal symmetry is broken by a SM
singlet, here we will use the $\mathbf{24}$ adjoint representation of
$SU(5)$.  We assign to the adjoint $\Sigma$ that breaks $SU(5)$ down
to the SM a horizontal charge $-1$ so that its vev
$\langle\Sigma\rangle = V_a$ with $V_a=V\> {\rm diag}\,(2,2,2,-3,-3)$
breaks also $U(1)_H$.  Here the normalization factor
$(2\sqrt{15})^{-1}$ of the $SU(5)$ generator $T_{24}$ has been
absorbed in $V$.  It is clear that the suppression factor $\epsilon =
V/M$ will now appear together with additional coefficients related to
the different entries in $V_a$.  We introduce also a second adjoint
$\overline{ \Sigma}$ with charge $+1$ in order to have a vectorlike
representation ($\Sigma\,, \overline \Sigma$) under the horizontal
symmetry, and allow for GUT scale masses.  With this modification, the
heavy FN fields responsible for inducing the mass operators of the
light fermions are no more restricted to lie only in the $\mathbf{5}$
or $\mathbf{10}$ representations of the group, as is the case when the
$U(1)_H$ breaking is triggered by a singlet.  Higher dimensional
representations like $\mathbf{15}$, $\mathbf{45}$, $\mathbf{70}\dots$
are now allowed, with the only restriction that the relevant vertices
involved in the construction of the effective operators must be
invariant under $SU(5)\times U(1)_H$.

\begin{figure}[t]
 \begin{center}
  \begin{pspicture}(-4.5,.6)(4.5,3.1)
  \qline(-4.3,1)(-.6,1)
  \qline(-3.5,1)(-3.5,2.5)
  \qline(-1.5,1)(-1.5,2.5)
  \uput[u](-3.5,2.5){$\langle \mathbf{\bar{5}}^{\phi_d}\rangle$}
  \uput[u](-1.5,2.5){$\langle\Sigma\rangle$}
  \uput[u](-4,1){$\mathbf{\bar{5}}$}
  \uput[u](-3,1){$\mathbf{{10}}_{-1}$}
  \uput[u](-2,1){$\mathbf{\overline{10}}_{+1}$}
  \uput[u](-1,1){$\mathbf{10}$}
  \uput[d](-2.5,1){$M$}
  \ast(-2.5,1)  
  \uput{-.05}[u](-.15,1){+}
  \qline(.3,1)(3.8,1)
  \qline(1.2,1)(1.2,2.5)
  \qline(3,1)(3,2.5)
  \uput[u](1.2,2.5){$\langle\mathbf{\bar{5}}^{\phi_d}\rangle$}
  \uput[u](3,2.5){$\langle\Sigma\rangle$}
  \uput[u](.7,1){$\mathbf{\bar 5}$}
  \uput[u](1.7,1){$\mathbf{15}_{-1}$}
  \uput[u](2.6,1){$\mathbf{\overline{15}}_{+1}$}
  \uput[u](3.5,1){$\mathbf{10}$}
  \uput[d](2,1){$M$}
  \ast(2.1,1)
  \end{pspicture}
 \end{center}
\vspace{-.3cm}
 \caption{Diagrammatic representation of the contributions to the
          masses of the $\tau$ lepton and of the ${b}$ quark.  The
          subscripts $-1$ and $+1$ represent the horizontal charges of
          the FN fields.}
 \label{fig1}
\end{figure}
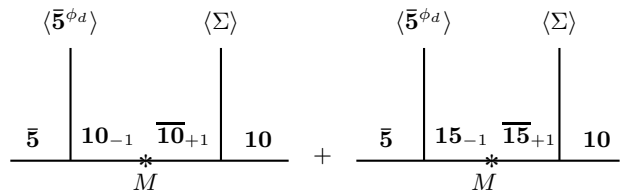

Let us now see what are the implications for the $b$ and $\tau$ mass
 operators that, being suppressed by only one power of $\epsilon$,
 require just one insertion of $\langle \Sigma\rangle$.  For the
 construction of these operators we assume one pair of vectorlike FN
 fields in the ten dimensional antisymmetric representations of
 $SU(5)$ $(\mathbf{10},\,\mathbf{\overline{10}})$ and a second pair in
 the symmetric $(\mathbf{15},\,\mathbf{\overline{15}})$.  Recalling
 the light fermions charge assignments $f+f_d=1$ and $t_3=0$,
 we need to assign a charge $-1$ to the $\mathbf{10}$ and
 $\mathbf{15}$, and $+1$ to the conjugate representations.  The light
 eigenvalues $m_b$ and $m_\tau$ of the resulting $3\times 3$ mass
 matrices can be computed with a very good approximation by summing up
 the contributions of the two  mass operators
 depicted in fig.~\ref{fig1}:
\begin{equation}\label{contributions}
{\cal L}_\pm = Y_{\pm}\ \mathbf{\bar 5}_a \>\langle \mathbf{\bar
  5}_b^{\phi_d}\rangle \>
\left(\frac{\delta^a_c\,\delta^b_d \, \pm\, 
\delta^a_d\,\delta^b_c}{2M}\right)\> \langle \Sigma^{d}_{\> f}\rangle\,
  \>\mathbf{10}^{fc}\,, 
\end{equation} 
where $a,b,\dots$ are $SU(5)$ indices. The term in parentheses
arises from integrating out the heavy FN fields, with the plus 
(minus) sign corresponding to the $\mathbf{15}$
($\mathbf{10}$), while $Y_\pm$ are two numbers of order unity.  By
projecting out the  vevs $\langle \mathbf{\bar
5}_b^{\phi_d}\rangle = v\> \delta_b^5$ and $\langle \Sigma^d_{\> f}\rangle=
V_a\> \delta^d_f$ we obtain: 
\begin{equation}\label{vacuum} 
{\cal L}_\pm = Y_{\pm}\>\frac{(-V_5\pm V_a)}{2M} \ \mathbf{\bar
5}_a\> 10^{a5}\> v\,, 
\end{equation} 
with $V_5\,$=$\,V_4\,$=$\,-3\,V$ and
$V_1\,$=$\,V_2\,$=$\,V_3\,$=$\,2\,V$.  Recalling the $SU(5)$
field assignments  $\mathbf{10}^{a5}=(d,\,d,\,d,\,e^c,0)^T$ and
$ \mathbf{\bar 5}_a=(d^c,\,d^c,\,d^c,\,e,\,-\nu)$, and summing up the
two contributions,  for the $b$ quark $(a=1,2,3)$ and $\tau$ lepton  
($a=4$) masses we obtain
\begin{eqnarray} 
m_b     &=& \frac{1}{2}\> (5\, Y_+ + Y_-)\> v\> \epsilon \\
m_\tau &=&   \frac{1}{2}\> (6 Y_-) \> v\> \epsilon\  =\ m_b + \delta m \\
\delta m &=&  \frac{5}{2}\> (Y_- - Y_+)\> v\> \epsilon.
\end{eqnarray} 
We see that mass unification ($\delta m=0$) is achieved only if the
two couplings $Y_\pm$ are equal. However, there is no reason for this
to happen.  $\delta m$ could be a positive or negative quantity, but
in general will be non vanishing.  The only requirement for $Y_\pm$ is
the same one that motivates the whole approach.  Namely, they must be
of order unity, implying that the hierarchy of the fermion masses and
mixing angles is solely determined by the horizontal symmetry, while
fluctuations of the dimensionless couplings around unity are only
responsible for their exact numerical values.  Since we do not have
yet a theory for the order one couplings, it is not possible to state
what is the size of these fluctuations, but it seems not an
unreasonable guess to assume an order of at least a few percent,
impling $\delta m/m\sim 10\%-20\%$.  From a bottom-up point of view,
it is clear that the low energy values of the $b$ and $\tau$ Yukawa
couplings when run up to the scale where the GUT symmetry and the
horizontal symmetry are broken will only approximately unify, even if
the unification of more fundamental couplings like $Y_+$ or $Y_-$ is
exact. We stress that $b$ and $\tau$ Yukawa non-unification is a
general outcome of breaking $U(1)_H$ through the adjoint  of
$SU(5)$. The particular FN representations introduced, only determine
the size of the deviations from unification.  For example, had we used
just one pair of $(\mathbf{5},\,\mathbf{\overline{5}})$ (and just one
coupling $Y_\mathbf{5}$) instead than $\mathbf{10}$ and $\mathbf{15}$,
we would have obtained $m_b/m_\tau = 2/3$. However, this deviates too
much from unity to be phenomenologically acceptable.


In this work we focused on the issue of $b$-$\tau$ Yukawa unification.
The prediction of an approximate but never exact unification is an
intriguing outcome of our framework. It would be interesting to
develop further the model and try to explain mass ratios like
$m_e/m_d$ and $m_\mu/m_s$ that, due to their large deviation from
unity, represent a real challenge to the idea of Gran Unification. We
plan to explore these issues in a future publication \cite{InPrep}.

\vspace{-3mm} 

%
\bigskip
\bigskip

\centerline{\bf Acknowledgments} \nobreak
 \medskip
The idea of breaking the horizontal symmetry with the adjoint of
 $SU(5)$ was suggested long ago to one of us (E.N.)  by Zurab
 Berezhiani.  We  acknowledge conversations with W. Ponce,
 D. Restrepo, A. Rossi and expecially with J. Mira.  This work was
 supported in part by Colciencias in Colombia.


\end{document}